\documentclass[prl,amsmath,amssymb,twocolumn,superscriptaddress]{revtex4}

\usepackage{amsmath,amssymb}
\usepackage[usenames]{color}
\usepackage{amssymb}
\usepackage{grffile}
\usepackage[pdftex]{graphicx}
\usepackage{amsmath, amstext, amssymb, amsfonts, amsxtra}
\usepackage{textcomp}
\usepackage{xspace}

\usepackage{color}

\begin{document}
\title{Disorderless quasi-localization of polar gases in one-dimensional lattices}
\author{W. Li}
\affiliation{Institut f\"ur Theoretische Physik, Leibniz Universit\"at Hannover, Appelstr. 2, 30167 Hannover, Germany}
\author{A. Dhar}
\affiliation{Institut f\"ur Theoretische Physik, Leibniz Universit\"at Hannover, Appelstr. 2, 30167 Hannover, Germany}
\author{X. Deng}
\affiliation{Institut f\"ur Theoretische Physik, Leibniz Universit\"at Hannover, Appelstr. 2, 30167 Hannover, Germany}
\author{K. Kasamatsu}
\affiliation{Department of Physics, Kindai University, Higashi-Osaka 577-8502, Japan}
\author{L. Barbiero}
\affiliation{Center for Nonlinear Phenomena and Complex Systems, Universit\'e Libre de Bruxelles, CP 231, Campus Plaine, B-1050 Brussels, Belgium}
\author{L. Santos}
\affiliation{Institut f\"ur Theoretische Physik, Leibniz Universit\"at Hannover, Appelstr. 2, 30167 Hannover, Germany}

\date{\today}

\begin{abstract}
One-dimensional polar gases in deep optical lattices present a severely constrained dynamics due to the 
interplay between dipolar interactions, energy conservation, and finite bandwidth. 
The appearance of dynamically-bound nearest-neighbor dimers enhances the role of the $1/r^3$ dipolar tail, 
resulting, in the absence of external disorder, in quasi-localization via dimer clustering for very low densities and moderate dipole strengths. 
Furthermore, even weak dipoles allow for the formation of self-bound superfluid lattice droplets with a finite doping of mobile, but confined, holons. 
Our results, which can be extrapolated to other power-law interactions, 
are directly relevant for current and future lattice experiments with magnetic atoms and polar molecules.
 \end{abstract}

\maketitle


Recent years have witnessed a major interest on the dynamics of isolated many-body quantum systems~\cite{Polkovnikov2011,DAlessio2016,Borgonovi2016,Basko2006,Nandkishore2015}. This interest has been largely triggered by impressive experimental 
developments, especially in cold gases~\cite{Bloch2008} and trapped ions~\cite{Blatt2012}, which realize almost perfect isolation~\cite{Kinoshita2006,Gring2012,Richerme2014,Jurcevic2014,Gaerttner2017}.  
Particular attention has been paid to atom dynamics in deep optical lattices, as in seminal experiments on single-particle and 
many-body localization in the presence of disorder~\cite{Billy2008,Roati2008,Deissler2010,Kondov2011,Jendrzejewski2012,Schreiber2015}.
However, in addition to energy conservation, tight-binding dynamics in deep lattices is largely determined by the finite bandwidth. This leads to the dynamical formation of 
(meta)stable states. A prominent example 
is provided by the so-called repulsively-bound pair, a pair of particles in the same lattice site that, although thermodynamically unstable,  remains 
dynamically bound if the interaction strength exceeds the lattice bandwidth~\cite{Winkler2006,Strohmaier2010}. The presence of repulsively-bound pairs leads, even for weak interactions, 
to a strong slow-down of the lattice dynamics~\cite{Schneider2012,Ronzheimer2013}.

Whereas experiments with contact-interacting particles realize Hubbard models with only on-site interactions, 
extended Hubbard models~(EHMs) with inter-site interactions may be realized using particles that interact via power-law potentials. This is the case 
of Rydberg atoms, with strong van-der-Waals interaction at nearest-neighbors~\cite{Saffman2010,Browaeys2016}, and of polar lattice gases with strong 
dipole-dipole interactions~(DDI), in particular magnetic atoms and polar molecules. Inter-site spin-exchange has been observed using chromium~\cite{DePaz2013} and KRb~\cite{Yan2013}, whereas an EHM with nearest-neighbor interactions has been realized using erbium~\cite{Baier2016}. Although EHM experiments with polar molecules remain a challenge 
due to inelastic losses~\cite{Ospelkaus2010, Mayle2013}, the latter may be avoided by using fermionic molecules~\cite{DeMarco2018}. 
In addition to leading to new ground-state physics~\cite{Lahaye2009,Baranov2012}, 
strong DDI change radically the lattice dynamics. 
Inter-site interactions, even just between nearest-neighbors, lead to 
non-local repulsively-bound pairs~\cite{Nguenang2009} and clusters at different sites, which significantly slow down the dynamics~\cite{Barbiero2015}.



\begin{figure}[t]
\begin{center}
\includegraphics[clip=true,width =0.75\columnwidth]{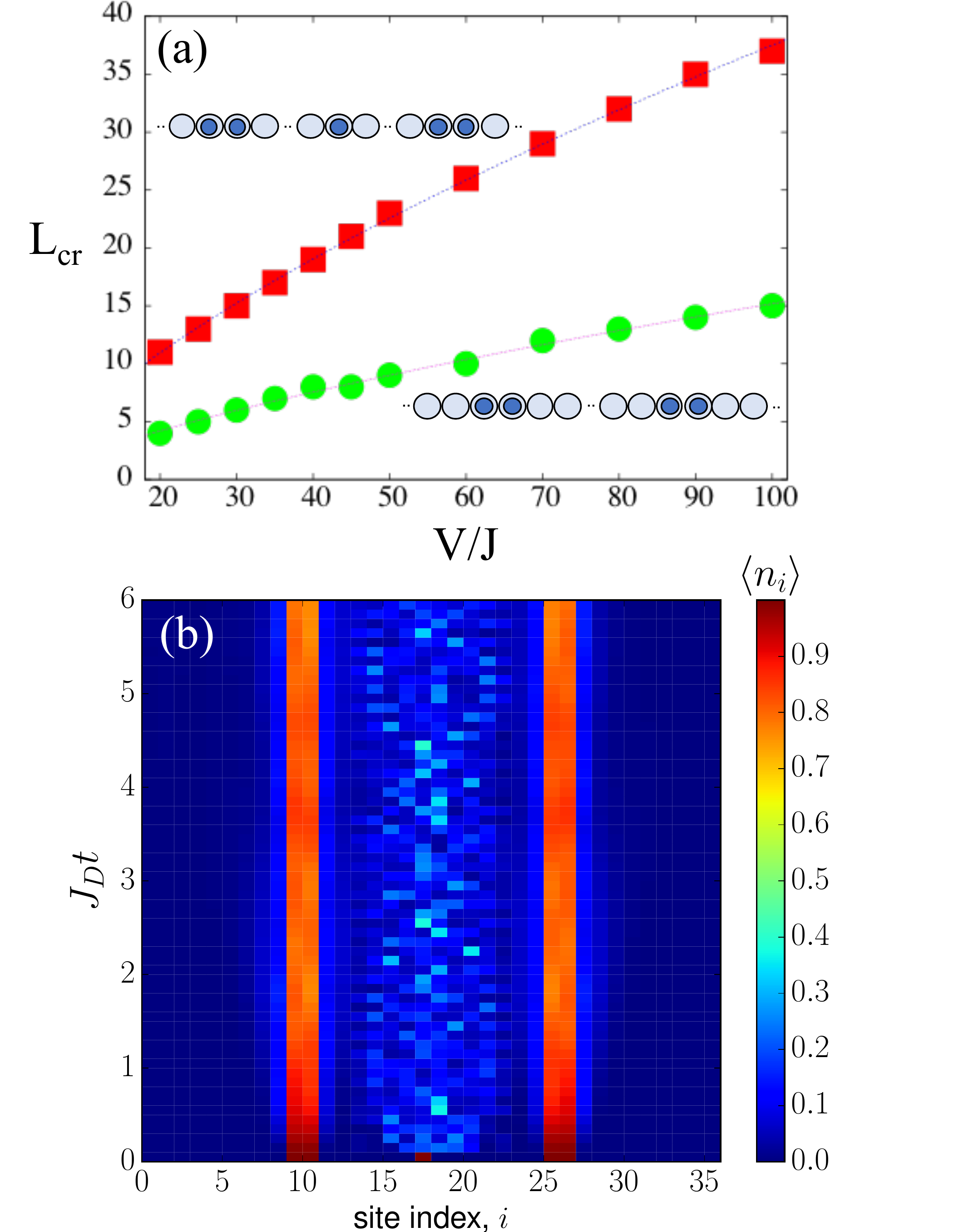}
\caption{(Color online) (a) Squares~(circles) indicate  $L_{cr}$~(see text) for two dimers with~(without) a singlon in between.  In both cases $L_{cr}\propto (V/J)^{2/3}$~(dotted curves).
(b) Time-dependent density-matrix renormalization group~(t-DMRG) results~\cite{Wall2012} of  $\langle \hat n_j \rangle(t)$ for $V/J=50$ for two dimers initially $15$ sites apart 
and an intermediate singlon. Whereas the singlon quickly delocalizes in the inter-dimer space, the NNDs remain at fixed distance for $t\gg 1/J_D$ 
despite the large separation.}
\vspace{-0.5cm}
\label{fig:1}
\end{center}
\end{figure}



\begin{figure*}[t]
\begin{center}
\vspace{-0.4cm}
\includegraphics[clip=true,width =2\columnwidth]{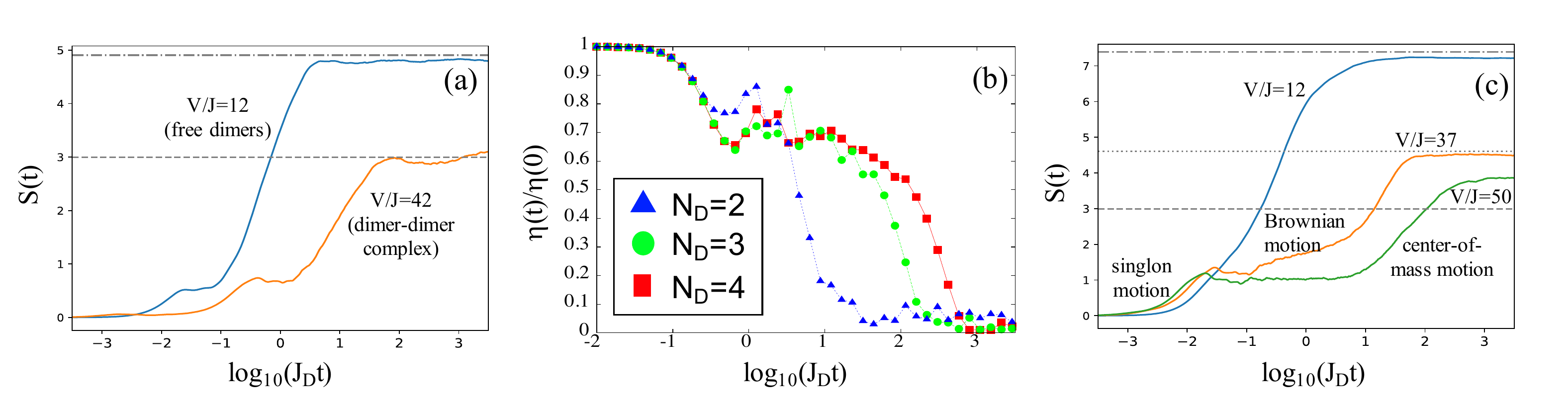}
\vspace{-0.2cm}
\caption{(Color online) (a) Shannon entropy $S(t)$, evaluated using exact diagonalization for $25$ sites and periodic boundary conditions for two NNDs initially $5$ sites apart, for 
$V/J=12$~(free dimers, in blue) and $42$~(bound dimes, in orange). Horizontal lines indicate $S_{max}$ for unbound~(dashed-dotted) and bound~(dashed) NNDs. 
(b) Inhomogeneity $\eta(t)/\eta(0)$ as a function of $J_D t$ for $V/J=40$ for clusters of 
$N_D=2$~(triangles), $3$~(circles), and $4$~(squares) NNDs, initially with three sites in between each NND in 
a lattice with $5(N_D+1)$ sites and periodic boundary conditions~(particle filling $\simeq 0.3$).  The homogeneization time~(which occurs for $\eta\simeq 0$) increases exponentially 
with the addition of each additional NND to the cluster.
(c) Same as (a) but for two NNDs initially $7$ sites apart and a singlon in between for $V/J=12$~(blue), $37$~(orange), and $50$~(green). 
Horizontal lines indicate $S_{max}$ for a free dimer-singlon gas~(dashed-dotted), 
and for dimers at a fixed distance with a singlon freely moving in between them~(dotted) and pinned fixed in between them~(dashed).}
\vspace{-0.5cm}
\label{fig:2}
\end{center}
\end{figure*}

In this Letter, we show that the formation of dynamically-bound dimers leads, in absence of disorder, to quasi-localization due to dimer clustering 
for surprisingly low densities and moderate dipole strengths. Moreover, in-lattice expansion experiments~\cite{Schneider2012,Ronzheimer2013} can reveal, 
even for weak dipoles, the formation of superfluid self-bound lattice droplets~\cite{footnote-SF-Metal}. Our results, which can be directly extrapolated to other power-law interactions, 
are directly relevant for current and future experiments on magnetic atoms and polar molecules.


\paragraph {Model.-} We consider polar bosons in a 1D lattice, which for simplicity are assumed as hard-core particles, i.e. 
the on-site interaction is so strong that maximally one boson is allowed per site. The system is described by the extended Bose-Hubbard Hamiltonian~(EBHM):
\begin{equation}
\hat H= -J\sum_j \left ( \hat a_j^\dag \hat a_{j+1} +\mathrm{H. c.} \right ) + \frac{V}{2} \sum_{i\neq j} \frac{1}{|i-j|^3} \hat n_i \hat n_j ,
\label{eq:H}
\end{equation}
with $\hat a_j$~($\hat a_j^\dag$) the annihilation~(creation) operator for bosons at site $j$, $\hat n_j=\hat a_j^\dag \hat a_j$, $(\hat a_j^\dag)^2=0$, 
$J$ the hopping rate, and $V$ the DDI strength between nearest-neighbors. Our results should not change qualitatively for single-component fermions.

\paragraph{Dynamically-bound dimers.-}  Large-enough $V/J$ leads to quasi-conservation of the number of nearest-neighbor links,  resulting in dynamically-bound nearest-neighbor dimers~(NNDs). 
For $V/J>7$, a NND can be considered an eigenstate~(the tightest two-particle bound eigenstate is a NND with a probability $>90\%$~\cite{footnote-SM}). 
We first consider a gas formed only of NNDs, which can be prepared e.g. using superlattices~\cite{Schreiber2015}. 
Dimers move by two consecutive hops, with effective hopping rate $J_D=8J^2/7V$. Their dynamics is approximately given by the EBHM
\begin{equation}
\frac{\hat H_D}{J_D}\! =\!-\!\sum_l \left ( \hat D_l^\dag \hat D_{l+1} +\mathrm{H. c. } \right ) + \frac{V}{J_D}\! \sum_{l, L\ge 1}\! f(L) \hat N_l \hat N_{l+L+2},
\label{eq:HD}
\end{equation}
where $\hat D_l^\dag=\hat a_l^\dag \hat a_{l+1}^\dag$ creates an NND at sites $l$ and $l+1$, $\hat N_l=\hat D_l^\dag \hat D_l$, and 
 $f(L)=\left (2 (L+2)^{-3}+(L+1)^{-3}+(L+3)^{-3} \right )$ characterizes the DDI between two dimers separated by $L$ sites.  
 Using $\hat H_D$ we may determine the critical number of empty sites between the dimers, $L_{cr}$, such that if 
 the initial inter-dimer separation $L_0<L_{cr}$, then at any other time $t$, this separation remains well 
 fixed, which we quantify by imposing that the variance of the inter-dimer distance $\Delta L < \sqrt{L_0}$ for $J_D t=100$~\cite{footnote-SM}. 
As expected from a simple inspection of $\hat H_D$, $L_{cr}\propto (V/J)^{2/3}$~(Fig.~\ref{fig:1}(a)). 
E.g., for $V/J=50$, two dimers $L_0=7$ sites apart remain at a tightly fixed relative distance, forming a dimer cluster despite their weak mutual dipolar interaction~($V f(L_0)\simeq 0.5 J$). 

The formation of dimer clusters leads to a strongly slowed-down dynamics. This is illustrated by the evolution of the Shannon entropy $S(t)=-\sum_{\{n_j\}} |c(\{n_j\},t)|^2 \log |c(\{n_j\},t)|^2$, 
obtained from the state of the system $|\psi(t)\rangle=\sum_{\{n_j\}} c(\{n_j\},t)|\{n_j\}\rangle$, with $|\{n_j\}\rangle$ the Fock states characterized by site occupations $n_j=0,1$. 
In Fig.~\ref{fig:2}(a) we depict our exact diagonalization results
obtained using Eq.~\eqref{eq:H}  for  $N_s=25$ sites with periodic boundary conditions, for two NNDs initially separated by $L_0=5$ sites, and 
$V/J=12$~($L_0>L_{cr}$) and $42$~($L_0<L_{cr}$). 
For  $t\ll 1/J_D$, $S(t)$ remains very low, since NNDs move via second-order hopping.  For $t\gtrsim 1/J_D$, the dimer cluster quickly unravels 
for $L_0>L_{cr}$, reaching a maximal entropy $S_{max}\simeq 2\ln N_s$. 
In contrast, for $L_0<L_{cr}$, a stable cluster of NNDs forms. As result $S(t)$ increases much slower, and only does it for  $t\gg 1/J_D$ 
due to the center-of-mass motion of the two dimers, up to a strongly reduced $S_{max}\simeq \ln N_s$.  

For sufficiently large densities, the presence of additional NNDs results in stable clusters of more than two dimers, which largely constrains entropy growth due to center-of-mass motion. 
This is illustrated in Fig.~\ref{fig:2}(b), where we depict for $V/J=40$, the evolution of the inhomogeneity parameter $\eta(t)=\sum_j |\langle \hat n_j \rangle -N/L |^2$~($\eta\simeq 0$ indicates homogeneization), 
obtained using exact diagonalization of $\hat H_D$ for $N_D=2$, $3$ and $4$ dimers initially separated by three empty sites in a lattice with $5(N_D+1)$ sites (particle filling $\simeq 0.3$ in all cases). Note that the homogeneization time increases by one order of magnitude with every dimer added to the cluster. 
Polar NNDs have hence a much stronger effect on the lattice dynamics than non-polar on-site repulsively-bound pairs~\cite{Petrosyan2007}. Contrary to the latter, where the larger 
effective mass of the pairs just leads to a slow-down, in polar gases weak dimer hopping is out-competed by the dipolar tail even at large distances, leading to quasi-localization via clustering even for dilute dimer gases and moderate dipoles.


\paragraph{Brownian motion.-} The presence of singlons~(i.e. isolated particles) between the NNDs radically changes the dynamics. 
On one hand, for weak-enough dipoles a singlon and a NND can approach at one site of distance, and then 
resonantly swap their positions, resulting in an effective dimer recoil. These recoils induce a Brownian-like 
dimer motion for $t\lesssim 1/J_D$.  In Fig.~\ref{fig:2}(c) we depict $S(t)$ for $V/J=12$, $37$ and $50$ for the case of one singlon initially 
in between two NNDs separated by $7$ sites~\cite{footnote-BM}. 
For $t\lesssim 1/J$, $S(t)$ grows linearly due to singlon motion between NNDs. For $1/J\lesssim t \lesssim 1/J_D$,  
Brownian motion results in an increase of $S(t)$, distinctly visible for $V/J=37$, 
which is then sped up by correlated dimer hopping for $t\gtrsim 1/J_D$. Brownian motion
is absent in the singlon-free NND gas~(Fig.~\ref{fig:2}(a)), and for large $V/J=50$~(Fig.~\ref{fig:2}(c)) for which singlons and NNDs cannot approach at one site from each other.


\paragraph{Singlon-gluing.-} Large-enough $V/J$ results, on the other hand, in a dramatic singlon-induced enhancement of the inter-dimer binding.
Due to the DDI, a singlon between two NNDs experiences a box-like potential~\cite{footnote-SM}, freely moving up to a distance $r_B$ from the dimers, with $V/r_B^3\sim J$, fully delocalizing 
within a time scale $1/J$ over the box length $L$~(Fig.~\ref{fig:1}(b)). Due to the singlon-dimer interaction, dimer motion changes the energy of the confined singlon. Hence the dimer dynamics 
is not only constrained, as above, by inter-dimer interaction, but also (and dominantly) by the change in the singlon energy. 
This mechanism resembles that discussed, for non-polar gases, in Refs.~\cite{Grover2014,Schiulaz2015}, and also for polar gases in Ref.~\cite{Barbiero2015}, in which 
the interplay between slow and fast particles (in this cases NNDs and singlons) was shown to result in quasi many-body localization. However the surprisingly strong role of the DDI tail, 
which plays a crucial role in this paper, was overlooked in the simplified model of Ref.~\cite{Barbiero2015}, in which a cut-off of the DDI at next-to-nearest neighbors was considered. 

We have evaluated the gluing effect, by exactly solving a system of two NNDs with an intermediate singlon~\cite{footnote-SM}.
Our results~(Fig.~\ref{fig:1}(a)) confirm indeed that the critical inter-dimer distance $L_{cr}$ for dimer clustering, which remains proportional to $(V/J)^{2/3}$, is very significantly enhanced, 
i.e. intermediate singlons lead to a much more robust dimer-dimer binding. For $V/J=50$, two NNDs initially separated by $L_0=15$ sites remain bound at fixed distance
for $J_D t\gg 1$~(Fig.~\ref{fig:1}(b)), despite the extremely small inter-dimer interaction $Vf(L_0)\simeq 0.02J$.

Singlon-gluing crucially affects the lattice dynamics of even very dilute 1D polar lattice gases for moderate dipoles.
A lattice gas with filling $\rho\ll 1$ is formed mainly by singlons, with an additional small density $\rho_D=\rho^2$ of NNDs~\cite{footnote-NNND}. 
Hence, for a sufficiently large $V/J$ that precludes Brownian motion, 
singlon-gluing leads to dimer clustering for $\rho\gtrsim \rho_{cr}$, with $\rho_{cr}^2 \simeq 1/ L_{cr}$.
As for the case of the singlon-free dimer gas, larger clusters of more than two NNDs prevent the center-of-mass motion that results in the entropy growth 
at long times observed in Fig.~\ref{fig:2}(c). Hence even moderate DDI result for very low densities~(for $V/J=50$, $\rho_{cr}\simeq 0.2$) into a  
quasi-localization via massive singlon-induced dimer clustering. Note that this estimation is conservative~\cite{footnote-NNND}, since for lower $\rho$, even in absence of massive clustering, the formation of clusters with few NNDs already severely constrains the dynamics within finite experimental lifetimes. 

\begin{figure}[t]
\begin{center}
\vspace*{-0.5cm}
\includegraphics[clip=true,width =0.7\columnwidth]{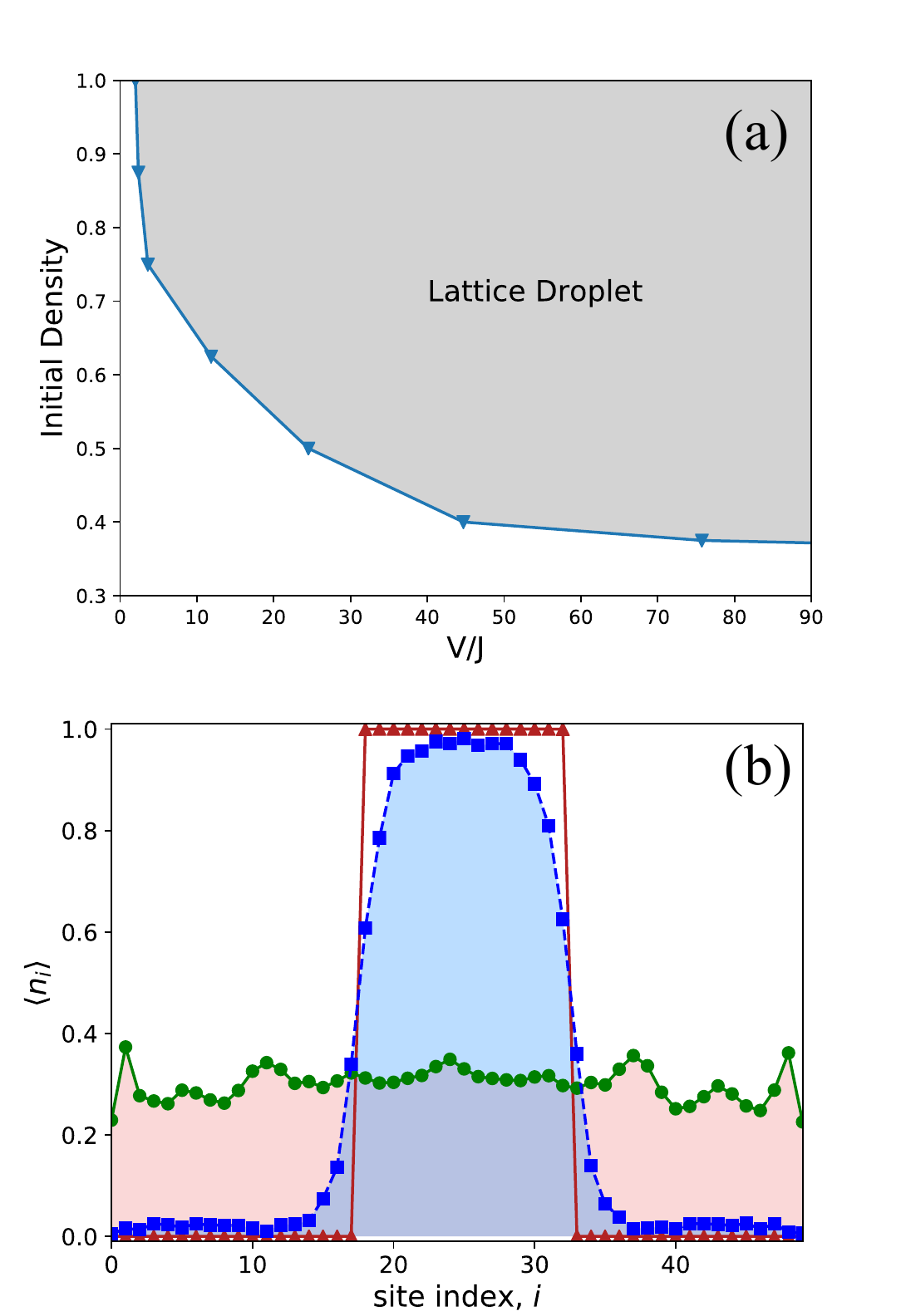}
\caption{(Color online) (a) $(V/J)_{cr}(\rho)$ for self-bound droplets obtained using exact diagonalization calculations in $16$ sites. 
The particles are initially created in the ground state (with $V=0$) of a box trap within the central $8$ sites. 
We determine $F(V/J,\rho)=\xi(t_f)/\xi(0)$, where $\xi(t)=\rho_c(t)-\rho_{av}$, with $\rho_{av}$ the density for a fully homogeneous gas in the lattice, and $\rho_c(t)$ the central density and  
$t_f=100t_D$, with $t_D$ the homogenization time for $V=0$. We determine $(V/J)_{cr}$ as that for which $F((V/J)_{cr},\rho)=0.1$. 
(b) Density distribution, obtained using t-DMRG~\cite{Wall2012}, for a gas initially confined with $\rho=1$~(red triangles) at $Jt=30$ after release, for $V/J=1$~(unbound, green circles) and 
$V/J=2.5$~(lattice droplet, blue squares).}
\vspace{-0.5cm}
\label{fig:3}
\end{center}
\end{figure}

\paragraph{Lattice droplets-} Even much weaker DDI may dramatically impact the dynamics, as illustrated by the analysis of in-lattice 
expansion. We consider a lattice gas of filling $\rho\leq 1$ initially prepared, with $V=0$, in the ground-state of a box-like potential~\cite{footnote-box}.
At time $t=0$ the box trap is released and $V>0$ is applied~(note that $V$ can be controlled by means of the dipole orientation).
In contrast to non-polar experiments~\cite{Schneider2012,Ronzheimer2013}, where stable or partially-stable on-site repulsively-bound pairs still allowed for an overall~(slowed-down) expansion, in the polar case 
there is a critical ratio $(V/J)_{cr}(\rho)$ such that the initial cloud remains self-bound~(Fig.~\ref{fig:3}(a)). These dynamically self-bound lattice droplets present a finite final average $\rho'<1$~(Fig.~\ref{fig:3}(b)), i.e.  holons~(empty sites) remain mobile but confined within a droplet. As a result, lattice droplets remain superfluid.
Note that for $\rho=1$, droplets occur already for $V/J\simeq 2.5$, which is readily available in current erbium experiments~\cite{Baier2016}. 
For finite holon doping, $(V/J)_{cr}$ increases~(Fig.~\ref{fig:3}(a)), but even rather dilute droplets 
may form for a moderate $V/J$.



\begin{figure}[t]
\begin{center}
\vspace*{-0.5cm}
\includegraphics[clip=true,width =0.7\columnwidth]{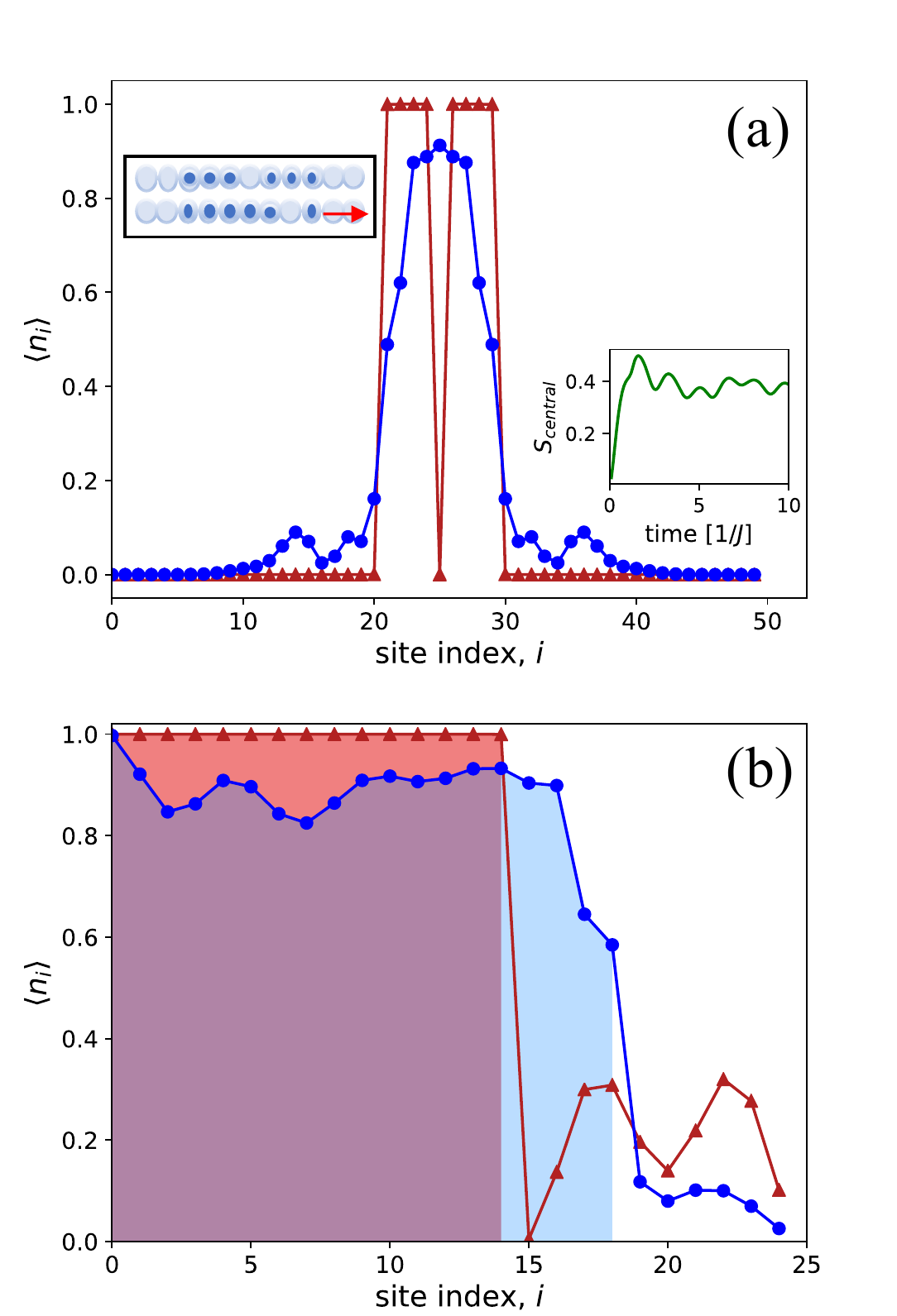}
\caption{(Color online) (a) Density distribution for a droplet with a holon initially in the middle~(red triangles) after a time $Jt=6$~(blue circles), for $V/J=30$. Partial holon evaporation results in particle ejection~(left inset). However 
evaporation is inefficient, as shown by the particle-hole entropy (averaged over the five central sites) depicted in the right inset. 
(b) Initial density distribution with a droplet with $\rho=1$ and two singlons outside~(red triangles) and distribution after $Jt=55$~(blue circles). The shadowed region 
is that of the droplet. Note the singlon aggregation at the droplet edge. The figures have been obtained using t-DMRG calculations~\cite{Wall2012}.}
\vspace{-0.5cm}
\label{fig:4}
\end{center}
\end{figure}


Holon confinement within a droplet results for large-enough $V/J$ from the effective trapping induced by the boundaries of the droplet through the DDI tail. 
For $V/J\lesssim 8$, this mechanism is insufficient, since only nearest-neighbor DDI are relevant for the holon dynamics. This is illustrated in Fig.~\ref{fig:4}(a), where we analyze 
a holon initially at the center of a droplet with $\rho=1$. The holon expands by resonant nearest-neighbor hops all the way to the droplet edges. 
When the latter occurs, the last particle of the droplet may escape without breaking any nearest-neighbor bond~(left inset of Fig.~\ref{fig:4}(a)). 
Although this introduces a mechanism for reducing the particle-hole entropy of the droplet, holon evaporation, which only occurs at the droplet edges, 
becomes drastically inefficient for growing droplet sizes since the holon quickly spreads uniformly within the droplet~(right inset of Fig.~\ref{fig:4}(a)).
As a result, holons remain confined within the droplet.
Interestingly, the converse of holon evaporation is also possible, i.e. a singlon initially outside the droplet may get attached to its edge, pushing a holon inside the droplet. Since holons 
remain trapped, this leads to singlon aggregation to the droplet with the consequent reduction of the droplet density~(Fig.~\ref{fig:4}(b)).
Evaporation and aggregation eventually equilibrate, such that for droplets placed in a singlon environment the 
outside singlon density equals the inside holon doping.

\paragraph{Conclusions.-} One-dimensional polar gases in deep lattices present a severely constrained dynamics. The formation of dynamically-bound dimers dramatically enhances the 
role of the $1/r^3$ dipolar tail, leading to quasi-localization in absence of disorder via the clustering of dynamically-bound dimers even for very small densities and 
moderate dipole moments. Moreover, polar lattice gases may form, even for very weak dipoles, self-bound superfluid lattice droplets 
with a finite doping of confined but mobile holons. Our results hint to inherent difficulties associated to particle-hole entropy removal in polar lattice gases, 
which must be carefully addressed for e.g. the creation of ground-state phases.  

Our work is directly relevant for lanthanide experiments. For example, for $^{164}$Dy in a deep UV lattice with inter-site separation $l=180$nm,  and orienting the dipoles along the lattice, 
$|V|/J\simeq 0.02 \eta^{-3/4} e^{-2\sqrt{\eta}}$, with $\eta$ the lattice depth in recoil units. For $\eta=23$, $V/J\simeq 30$, with a hopping rate $J/\hbar\simeq 93$s$^{-1}$. The corresponding 
dimer-hopping time is $1/J_D\simeq 280$ms, and hence the effects of dimer clustering may be readily probed in few seconds, well within typical experimental lifetimes. 
Our results are also applicable to future experiments with polar molecules, where even larger $V/J$ ratios may be achieved, and 
may be easily extrapolated to systems with other power-law interactions.

\acknowledgements
We thank L. Chomaz, F. Ferlaino, C. Menotti, M. Mark, T. Pfau, and A. Recati for discussions. 
W. L., X. D. and L. S. thank the support of the DFG (SFB 1227 DQ-mat and FOR2247). 
L. B. acknowledges ERC Starting Grant TopoCold for financial support.
K. K. thanks the support by KAKENHI (Grant No. 18K03472) from the JSPS.



\begin{thebibliography}{99}
\bibitem{footnote-SF-Metal} For Fermi gases the droplets would remain metallic.
\bibitem{footnote-BM} Due to periodic boundary conditions, a single singlon, which may present different winding numbers around the periodic system, will lead to 
Brownian motion of the NNDs, mimicking the behavior expected for NNDs in a singlon bath.
\bibitem{footnote-NNND} We consider neither nearest-neighbor clusters of $n>2$ particles, which occur with density $\rho^n$, nor the formation of dimers beyond nearest neighbors. 
For sufficiently large $V/J$ these dimers may become stable as well. Their presence, however, would result in an even stronger localization of the lattice gas.
\bibitem{footnote-box} More general confinements, e.g. harmonic traps, or finite-temperature effects do not modify our conclusions.
\end{thebibliography}
\end{document}


\title{Supplementary Material for "Disorderless quasi-localization of polar gases in one-dimensional lattices"}
\author{W. Li}
\affiliation{Institut f\"ur Theoretische Physik, Leibniz Universit\"at Hannover, Appelstr. 2, 30167 Hannover, Germany}
\author{A. Dhar}
\affiliation{Institut f\"ur Theoretische Physik, Leibniz Universit\"at Hannover, Appelstr. 2, 30167 Hannover, Germany}
\author{X. Deng}
\affiliation{Institut f\"ur Theoretische Physik, Leibniz Universit\"at Hannover, Appelstr. 2, 30167 Hannover, Germany}
\author{K. Kasamatsu}
\affiliation{Department of Physics, Kindai University, Higashi-Osaka 577-8502, Japan}
\author{L. Barbiero}
\affiliation{Center for Nonlinear Phenomena and Complex Systems, Universit\'e Libre de Bruxelles, CP 231, Campus Plaine, B-1050 Brussels, Belgium}
\author{L. Santos}
\affiliation{Institut f\"ur Theoretische Physik, Leibniz Universit\"at Hannover, Appelstr. 2, 30167 Hannover, Germany}

\date{\today}

\begin{abstract}
In this Supplementary material we provide further information about the analysis of the nearest-neighbor dimer stability, 
and the dimer-dimer and dimer-singlon-dimer complexes
\end{abstract}

\maketitle

\section{Dynamically-bound nearest-neighbor dimers}

The two-body problem of formation of dynamically-bound off-site dimers has been previously discussed in the literature, see e.g. Refs.~\cite{Nguenang2009, Barbiero2015}. 
Here we just briefly sketch the formalism. We denote as $|l,l+j\rangle$ the state with two polar particles placed at sites $l$ and $l+j$.
We introduce the Fourier transform: $|\psi_j(k)\rangle = \sum_l e^{ik(l+1/2)}|l,l+j\rangle$, with $k$ the center-of-mass momentum. The extended Bose-Hubbard Hamiltonian 
may be re-written for each $k$ as:
\begin{eqnarray}
\hat H(k)&=&-2J\cos (k/2) \sum_{j\ge 1} \left ( |\phi_j(k)\rangle\langle \phi_{j+1}(k)| +\mathrm{H. c.} \right ) \nonumber \\
&+&\sum_{j\ge 1} \frac{V}{j^3} |\phi_j(k)\rangle\langle \phi_j(k)|
\end{eqnarray}
Diagonalizing $\hat H(k)$ we obtain the eigenstates $|\psi_n(k)\rangle=\sum_j c_{j,n} |\phi_j(k)\rangle$ with eigenenergies $\xi_n(k)$. For a sufficiently large $V/J$ the spectrum 
presents two different types of states, scattering states (where the particles are not bound) and bound states. The latter states, irrespective of the sign of $V$, result 
in dynamically stable dimers. In our case, we assume $V>0$, and hence these states, as mentioned in the main text, can be considered the off-site equivalent of the well-known 
repulsively-bound pairs~\cite{Winkler2006,Strohmaier2010}. For low $V/J$ no bound-state occurs. At a sufficiently large $V/J$ a bound-state appears, which corresponds to a very good approximation 
to a nearest-neighbor dimer~(NND). For larger 
$V/J$ other bound-states, e.g. next-to-nearest-neighbor dimers, may occur as well. However, for our purposes only the stability of NNDs is relevant. 
Since the least-bound NNDs are those at $k=0$, we will focus on these dimers in the following.

For a sufficiently large $V/J$, the uppermost bound-state is basically an NND. By evaluating the overlapping, $P_D$, between $|\psi_1(k=0)\rangle$ and the 
tightest bound-state we may hence determine how tightly bound the NND is. We depict $P_D$ in Fig. 1. Note that for $V/J>7$ the tightest bound-state is 
over $90\%$ a NND, and hence an initial NND remains to a very good approximation an NND for the whole dynamics. The latter is the case in the following sections, where we assume 
a large-enough $V/J$.



\begin{figure}[t]
\begin{center}
\vspace*{0.5cm}
\includegraphics[clip=true,width =0.8\columnwidth]{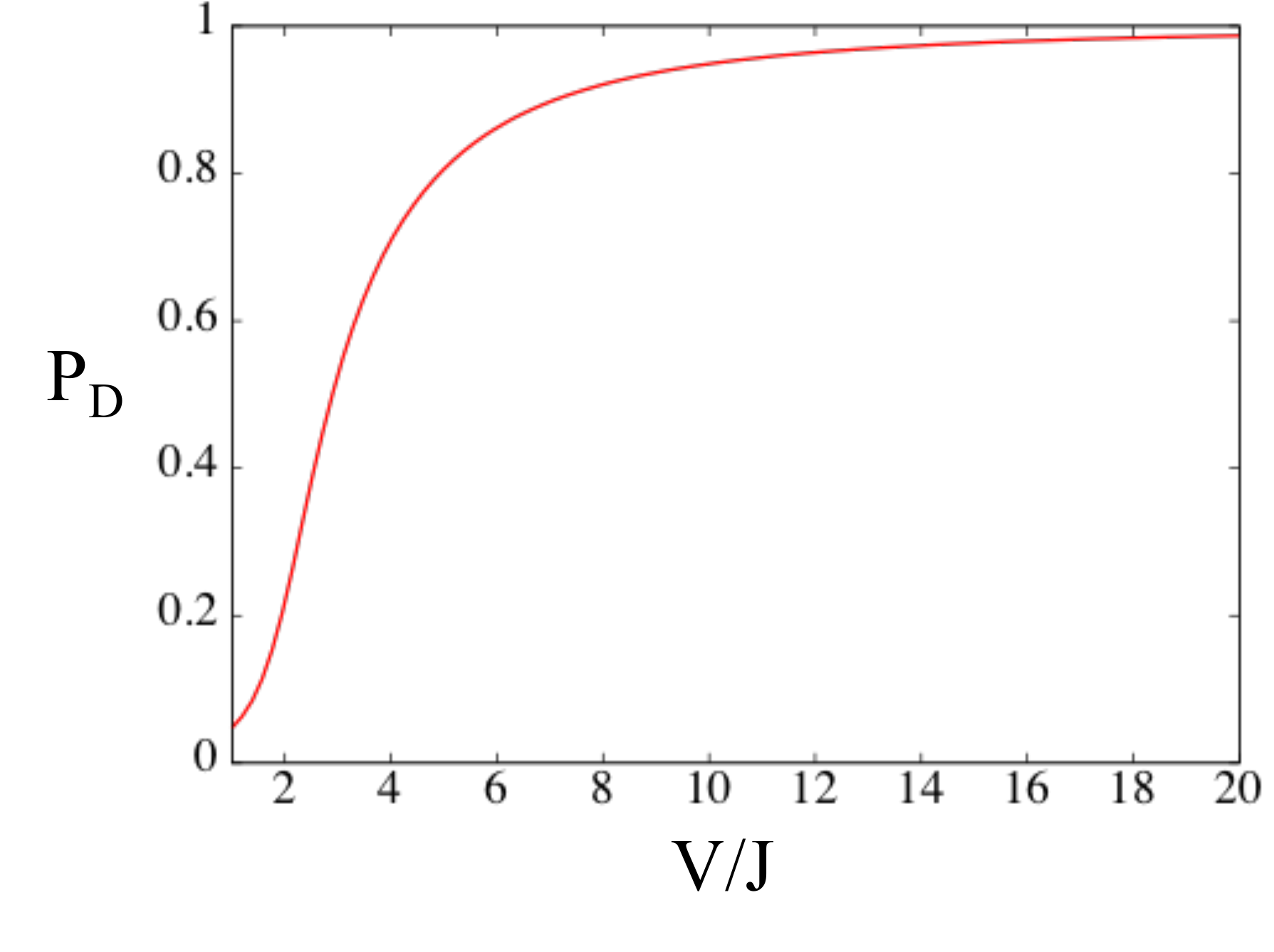}
\vspace*{-0.5cm}
\caption{(Color online) Probability $P_D$ that a NND is in the most tightly bound eigenstate. For the depicted values of $V/J$ 
there is only one bound state, and hence $1-P_D$ is the probability fraction that the NND unravels in time.}
\label{fig:S1}
\end{center}
\end{figure}


\section{Dimer-dimer model}

Assuming stable NNDs, we may evaluate the dimer-dimer clusterization by approximating the four-body dimer-dimer problem by a simplified two-body problem, using the Hamiltonian 
$\hat H_D$ introduced in Eq. (q) of the main text. We may then proceed in exactly the same way as in the previous section. For a center-of-mass momentum $k=0$ of the dimer-dimer 
complex, the Hamiltonian is:
\begin{eqnarray}
&&\!\!\!\frac{\hat H_D(k=0)}{J_D}=-2 \sum_{j\ge 1} \left ( |\tilde\phi_j\rangle\langle \tilde\phi_{j+1}|+\mathrm{H. c.} \right ) \nonumber \\
&&+ \frac{V}{J_D}\sum_{j\ge 1}  \left ( \frac{2}{j^3}+\frac{1}{(j+1)^3} +\frac{1}{(j-1)^3} \right )|\tilde\phi_j\rangle\langle \tilde\phi_j|,
\label{eq:HDk}
\end{eqnarray}
with $|\tilde\phi_j\rangle$ plays the same role for two tightly-bound NNDs at $|l,l+1\rangle$ and $| l+j,l+j+1\rangle$, as $|\phi_j(k=0)\rangle$ played for two particles separated by a distance $j$. 
Assuming an initial state $|\tilde\phi_{j_0}\rangle$ (two NNDs with $L_0=j_0-2$ empty sites in between), we may evaluate using $\hat H_D(k=0)$ the average dimer-dimer distance
$\langle L \rangle (t)$, and its variance $\Delta L(t)=\sqrt{\langle L^2\rangle -\langle L \rangle ^2}$. As a criterion for the stability of the dimer-dimer complex, we determine 
for different values of $V/J$ the critical $L_{cr}$, such that for $L_0<L_{cr}$ in the subsequent evolution, up to $J_D t=100$, $\Delta L < \sqrt{\langle L_0 \rangle}$, i.e. such that the 
inter-dimer distance remains well defined~(Fig.~1(a) of the main text).

\section{Dimer-singlon-dimer model}

In the following we introduce a model that permits a simple analysis of the singlon-gluing effect discussed in the main text. We consider two stable NNDs, and a singlon 
placed in between them. For simplicity, we do not consider the possibility of Brownian-like motion, i.e. of singlon-dimer swapping that may result in the escaping of the singlon from the inter-dimer region. 
This approximation is valid for large-enough $V/J$.  

The states of the dimer-singlon-dimer~(DSD) system are of the form $|\psi_n^{j,l}\rangle \equiv |D_l\rangle \otimes |\phi_n^{j,l} \rangle \otimes |D_{l+j}\rangle$, where $|D_l\rangle$ denotes an NND placed at sites $l$ and $l+1$. 
In absence of dimer motion, the interaction between the singlon and the two dimers results in an effective potential for the singlon:  
$V_S(l+2\leq x\leq l+j-1)=V\left (f_{DS}(x,l)+f_{DS}(x,l+j) \right )$, with $f_{DS}(x,l)=|x-l|^{-3}+|x-(l+1)|^{-3}$. The states $ |\phi_{n=1,\dots,j-2}^{j,l} \rangle$ above are the singlon eigenstates  in this 
effective box-like potential, with a corresponding singlon energy $\xi_n^j$.

We may then introduce the Fourier transform $\psi_n^j(k)\rangle =\sum_l e^{ik(l+(j+1)/2}|\psi_n^{j,l}\rangle$, with $k$ the center-of-mass momentum of the dimer-dimer pair. 
As above, we reduce our analysis to $k=0$, for which the dimer-dimer binding is minimal. We hence simplify  the notation $|\psi_n^j(k=0)\rangle \equiv |\psi_n^j\rangle$. 
Dimer hopping couples $|\psi_n^j\rangle$ with $|\psi_{n'}^{j\pm 1}\rangle$. The overall Hamiltonian for the DSD system is hence given by:
\begin{eqnarray}
\hat H_{DSD}&=&\sum_{j,n}\left ( \xi_n^j+V f(j) \right ) |\psi_n^j\rangle\langle \psi_n^k |  \nonumber \\
&-& \sum_{j,n,n'} \left [ \Lambda_{n,n}^j |\psi_{n'}^{j+1}\rangle\langle \psi_n^j | + \mathrm {H. c.} \right ],
\end{eqnarray}
where $V f(j)$ is the dimer-dimer interaction discussed in the main text~($L=j+2$), and the coupling between states of different inter-dimer distance is characterized by the overlappings
$\Lambda_{n,n'}^j=\sum_{s=2}^{j-1} \phi_n^j(s) \left [ \phi_{n'}^{j+1}(s+1)+\phi_{n'}^j(s) \right ]$, with $\phi_n^j(s)$ the amplitude of the state $|\phi_n^{j,l}\rangle$ in site $l+s$.

Assuming an initial condition with two dimers separated by $L_0$ sites, with the singlon initially in the middle of the effective box, we employ the Hamiltonian $\hat H_{DSD}$ to obtain (using 
exact diagonlaization) the DSD state at any other time. As for the dimer case we determine the inter-dimer separation, and define the critical initial separation $L_{cr}$ such that for $L_0<L_{cr}$, one has  
$\Delta L<\sqrt{L_0}$ for $J_D t=100$. In this way, we obtain the results of Fig. 1(a) of the main text.
